# Pan-African Citizen Science e-Lab: An Emerging Online Platform for Astronomy Research, Education and Outreach in Africa.

Miracle Chibuzor Marcel[1], Kassamba Abdel Aziz Diaby[2], Meryem Guennoun[3], Betty Rose Nabifo[4], Mohamed Elattar[5], Andoniaina Rajaonarivelo[6], Privatus Pius[7], Molly Nkamogelang Kgobathe[8], Immanuel Luis[9], Sigrid Shilunga[10], Nejmeddine Etteyeb[11], Keketso Qhomane[12], Samuel Nyangi[13], Tresford Chilufya Kalunga[14], Nunes Alfredo Assano[15], Edson Domingos Jequecene[16], MAFUKA LUSALA JOSEPH[17], Esaenwi Sudum[18], Jorbedom Leelabari Gerald[19], Christopher Tombe Louis Gore[20], Kareem Waleed Hosny[21], Nagat Yasser[22], Jocelyn Franck[23], MAMOUDOU KOUROUMA[24], BABOUCARR BOBB[25], Kebab Jaiteh[26], Salma Sylla[27], Hans ESSONE OBAME[28], Dennis Kiyeng[29], Thobekile Sandra Ngwanw[30], Tawanda Kelvin Simon[31], Saja Alhoush Sulayman[32], Salma Regaibi[33], Souley Yahaya[34], Tengwi Mogou Ornela[35], Henry Sanderson Viyuyi[36], Fortune Tatenda Matambo[37], Matthias Asare-Darko[38], Christian Kontoa Koussouwa Gbaba[39], Moisés da Silva[40], Ntahompagaze Joseph[41], Gilberto Gomes[42], Bongiwe Portia Mkhabela[43], Bauleni BVUMBWE[44], Tshombe Nkhowani[45], Mawugnon Axel Gahou[46], Sarah Abotsi-Masters[47], René Simbizi[48], Salomon Mugisha[49], Ahmed Saeed[50], Mohammed Yahya Alradi Eldaw[51], Allen Thomas[52], Ben Abdallah ridha[53], Dieumerci kaseha[54], Sherine Ahmed El Baradei[55], Nahla Hazem Hussein[56], BADO Fabrice[57], Ngozika Frances Anekwe[58], Arvind Ramessur[59], Mohamed Ali Koroma[60], Harold Safary[61], Oosthuizen Leonardo[62], Mdumiseni Wisdom Dabulizwe Dlamini[63], Mamadou Mahamat Djabbi[64], Nonofo Angela[65], Mamaja Jalloh[66], Mamadou Balde[67], Joy Olayiwola[68], Elijah Ibharalu[69], Thierry Martial TCHANGOLE[70], Kirubel Memberu[71], Lidia Dinsa[72], Chidozie Gospel Ezeakunne[73]

1. Pan-African Citizen Science e-Lab, FCT, Abuja: info@pacselab.space, miracle.c.marcel@gmail.com; 2. Université Félix Houphouët-Boigny; 3. Oukaimeden Observatory; 4. National Curriculum Development Centre; 5. IEEE New Beni Suef; 6. Haikintana Astronomy Association; 7. Department of Natural Science, Mbeya University of Science and Technology; 8. University of Botswana Astronomy Club; 9 & 10. Department of Physics, Chemistry and Material Science University of Namibia; 11. Tunisian Astronomical Society; 12. BlueCraneSpace Astronomy & Astrophysics Department - University of Pretoria; 13. Amateur Astronomical Society of Kenya; 14. Kabulonga Girls' Secondary School; 15 & 16. Detetives do Cosmos; 17. Astroclub Kongo Central; 18. The Astro Group of the Rivers State University; 19. The Astro Group of the Rivers State University; 20. Mayardit Academy for Space Sciences - University of Juba; 21 & 22. Pharaohs of Space; 23. University Marien Ngouabi; 24. IT Dreams and Promises of the University of Bamgui; 25. Physics Dept. University of Gambia; 26. Physics Dept. University of Gambia; 27. Orion Astrolab and Department of Physics, University Cheikh Anta Diop; 28. IAU - NAEC Gabon; 29. Space Partnerships and Applications Company Kenya; 30. Zimbabwe Astronomical Society; 31. Zimbabwe National Geospatial and Space Agency; 32. Amateur Astronomy Libya; 33. Steps Into Space Association; 34. Niger Space Surfer; 35. Astronomy Club of the University of Buea; 36. Zambia Space Explorers; 37. Zimbabwe National Geospatial and Space Agency; 38. PRAGSAC; 39. NGO Science Géologique pour un Développement Durable (SG2D); 40. Associação Angolana de Astronomia; 41. Physics Department, University of Rwanda, College of Science and Technology; 42. Angolan Space Program Management Office (GGPEN); 43. Galaxy Explorers; 44. Celestial Explorers; 45. Copperbelt University; 46. Sirius Astro-Club Benin; 47. Ghana Planetarium; 48 & 49. Physics Dept. University of Burundi; 50. Sudanese Asteroid Hunters; 51. Institute of Space Research and Aerospace (ISRA); 52. Center for Science Education; 53. Aljarid Astronomie; 54. Lubumbashi Astro Club; 55 & 56. Space -Water-Environment Nexus e - Center; 57. Laboratoire de Physique et de Chimie de l'Environnement, Université Joseph KI-ZERBO, Ouagadougou, Burkina Faso; 58. Phys. Dept. Chukwuemeka Odumegwu Ojukwu University Uli Campus Anambra State; 59. IAU - NOC Mauritius; 60. Cosmic Gazers Research Institute Sierra Leone; 61. Kenya Space Agency - Education and Outreach Students' Network; 62. Night Sky Tours; 63. Destiny Stars of the Univesity of Eswatini; 64. Toumaï; 65. Marang Junior Secondary School; 66. Sierra Leone Geospatial and Space Agency; 67. UNESCO Center for Peace USA in Guinea; 68. National Space Research and Development Agency, NASRDA HQ; 69. University of Benin; 70. CosmoLAB Hub Association; 71 & 72. Ethiopian Space Science Society; 73. University of Central Florida

**Abstract**

Citizen science offers an opportunity for ordinary people, known as citizen scientists or citizen astronomers in the context of astronomy, to contribute to scientific research. The Pan-African Citizen Science e-Lab (PACS e-Lab) was founded to promote and engage the African public in citizen science and soft astronomy research to advance space research and exploration and enhance space education and outreach. PACS e-Lab, in collaboration with several international astronomy research, education, and outreach organizations, currently runs several projects including but not limited to asteroid search, exoplanet photometry, research writing for peer-reviewed publications, astrophoto visual development, and Amateur Radio contact with astronauts aboard the International Space Station (ARISS). Despite several challenges, the group has engaged over 600 Africans from more than 40 countries and is working towards covering the entire continent in the future. PACS e-Lab's development efforts resonate with seven United Nations Sustainable Development Goals (UN-SDGs).

## 1.0 INTRODUCTION

Citizen science is the practice of public participation and collaboration in scientific research. This practice increases the scientific knowledge of amateur scientists. (Trouille et al, 2019). Through citizen science, volunteers, mostly non-professionals in the specified field, can contribute to various stages of scientific projects, such as data collection, analysis, and dissemination of results. Among all research fields, astronomy is at the top when considering data output, in other words, astronomy is a field of big data, most of which is freely accessible. Access to these datasets has made astronomy citizen science endeavors more pronounced (Schwamb, 2016). Astronomy citizen science over the years gained traction in developed countries especially its adoption in the educational sector, helping to inspire the next generation in Science, Technologies, Engineering, and Mathematics (Trouille et al, 2019; Pompea and Russo, 2020), while it has not gotten a deep root in developing countries in Africa.

The Pan-African Citizen Science e-Lab (PACS e-Lab) which is currently an online platform was formed to fill these needs, thereby making hands-on astronomy projects accessible to all Africans. Some of the objectives of the PACS e-Lab include the dissemination of citizen science and soft astronomy research projects that enable African participants to contribute to space research and exploration, providing the African participants with adequate training and educational resources to be able to engage in meaningful and effective citizen science and soft astronomy research projects, providing opportunities for our African participants to partner with internationally acclaimed space scientists and engineers in designing and conducting research projects, fostering collaborations with diverse national space agencies, industry, and academic institutions in Africa to support citizen science and soft astronomy research projects, informing the African scientific and public populations about the results of African citizen science programs and research etc. We aim to promote hands-on activities in astronomy & space science through citizen science and Soft Astronomy research in Africa as a means of advancing space exploration, and enhancing space education and outreach.

It should be noted that the PACS e-Lab as an organization does not create citizen science projects at the moment; rather, the organization collaborates with other organizations running citizen science initiatives mostly in the Global North to bring them to Africa. This practice bridges the knowledge gap in astronomy research, education and outreach between developed countries and developing countries in Africa.

This paper is organized as follows: Section 2 provides the methodology, which includes the organizational overview of PACS e-Lab, key projects, and how PACS e-Lab accesses astronomical resources. Section 3 showcases the results, including tables and charts from the survey. Section 4 discusses the results, including responses from the survey and the projects' contributions to the development of Africa. Finally, Section 5 presents the conclusion and a call to action.

2.0 **Methodology**

2.1 History and Founding:

The founding of PACS e-Lab is rather unconventional; the journey began during the COVID-19 lockdown with Miracle Chibuzor Marcel. On May 30, 2020, NASA and SpaceX conducted a historic mission, launching the Crew Dragon spacecraft atop a Falcon 9 rocket. This was the first crewed launch from American soil since the Space Shuttle program ended in 2011 and the first time a private company (SpaceX) transported astronauts to the International Space Station (ISS). Before this event, a special Facebook group called the Launch America NASA Social Group was created by NASA, where people worldwide joined to celebrate the event. Through this group, Miracle Chibuzor Marcel connected with several space industry professionals linked to astronomy education and outreach, particularly with the International Astronomical Search Collaboration (IASC) (Miller, 2018) and Slooh web telescopes (Gershun et al, 2014).

He became a registered participant of IASC and participated in the asteroid search from November 9 to December 4, 2020. During his participation, he realized that only a few individuals and groups from a handful of African countries were involved in this project. Inspired by this, he decided to spread the citizen science project across the continent, engaging both industry professionals and non-professionals in Africa.

At that time, most of the groups Miracle Chibuzor Marcel engaged with were registered under the International Asteroid Search Campaign, which caters to teams not affiliated with any regional group. However, as the number of African groups grew, Mrs. Davis Cassidy, the IASC coordinator, suggested creating an unofficial African regional asteroid search group called the Pan-African Asteroid Search Campaign. This group later became official after the African Astronomical Society, through Dr. Charles Takalana, was informed about the group's activities and sent a letter of support to IASC to formalize the Pan-African Asteroid Search Campaign. This happened in November 2022.

In the latter years, Miracle Chibuzor Marcel expanded the available hands-on activities by collaborating with more astronomy education and outreach organizations, primarily in the Global North, to include a greater variety of interactive experiences. These initiatives are described in detail in Section 3.

However, as the number of projects increased, the name Pan-Africa Citizen Science e-Lab (PACS e-Lab) was adopted to describe an online laboratory for disseminating citizen science projects in Africa. On June 1, 2023, PACS e-Lab became a registered non-profit organization in Nigeria, the certificate of registration is shown in Figure 1. PACS e-Lab's official website is [www.pacselab.space](www.pacselab.space) and links to all social media platforms are included therein, the official logo is also shown in Figure 1.

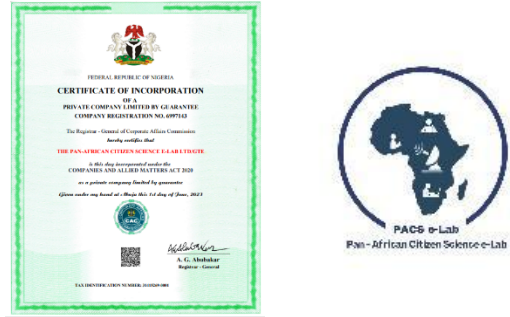

*Figure 1: Registration Certificate and official logo of PACS e-Lab. The logo was designed voluntarily by https://linktr.ee/Emmanuel_Adeoluwa_Ajala*

2.2 Structure and Team Formation

After Miracle Chibuzor Marcel's first participation in the IASC asteroid search, he devised a plan to engage more Africans in the project. His initial approach involved reaching out to astronomy groups across Africa via social media platforms like Facebook and LinkedIn, Google searches, emails, and referrals from individuals already involved in the project. Most of these groups responded positively, leading to introductory and training online meetings.

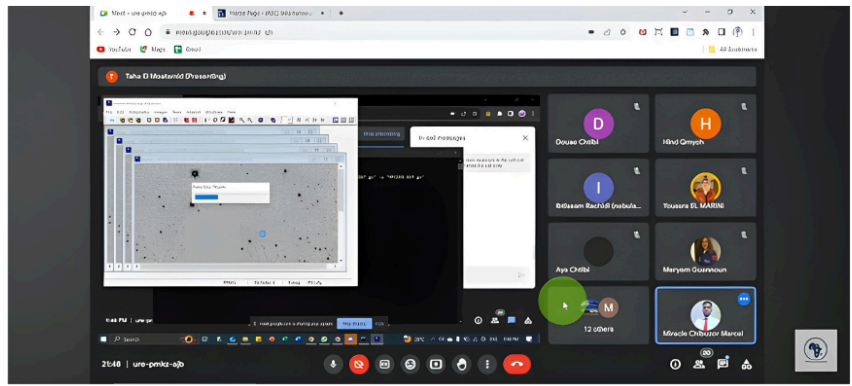

*Figure 2: One of many PACSe - Lab's online training sessions with teams across the continent, each lasting over three hours*

Recently, PASC e-lab management improved its recruitment methodology to boost recruitment by partnering with organizations that welcome participants from Africa, such as the International Astronomy and Astrophysics Competition (IAAC), which has a database of over 900 participants from Africa. In 2023, PACS e-Lab supported 10 African participants from 10 different countries by covering their IAAC participation fees. This effort was recognized by them, and PACS e-Lab became one of their partners, with a promise to promote its endeavors among African participants in their database.

All PACS e-Lab projects are suitable for all individuals. The only requirements are having an adequate internet connection, a computer with a Windows operating system, and a passion for astronomy. The organization welcomes everyone, amateurs and professionals alike, in Africa who wish to join, regardless of age, religion, country, sexual orientation, academic or professional affiliation and levels, etc.

When PACS e-Lab management sends invitation letters out, the focus is mainly paid to entities with access to the members of local communities like the space education and outreach departments of national space agencies, STEM departments of African universities, astronomy organizations, secondary schools, and astronomy groups. These entities typically have existing outreach programs, some resources to execute projects, and access to the public. PACS e-Lab through the general WhatsApp group and other communication means liaises with the leaders of these groups through training and information sharing, who then disseminate the information to their members.

Additionally, PACS e-Lab has some volunteering members who maintain the group's database, create graphic designs, recreate tutorial videos in their local languages including French and Arabic, and train teams who do not speak English. Despite various challenges, its endeavors have become some of the most significant astronomy initiatives on the continent, connecting teachers, students, and space enthusiasts across Africa, resonating with points raised by Pović et al, 2018 on activities in astronomy and space sciences (A&SS) in Africa.

Ever since December 4th, 2020 the official date for the founding of PACS e-Lab, the group has recruited and engaged over 600 individuals from 45 African countries, including both North Africa and Sub-Saharan Africa, and has continued to expand. The goal of the PACS e-Lab is to cover the entire African continent. The group's endeavors have sparked a rise in citizen science participation among African youths, leading to the formation of new amateur astronomy groups and the strengthening of existing ones. Recently, the group started welcoming collaborators from outside Africa who are interested in adopting their projects.

## 2.3 Key Projects

PACS e-Lab projects are classified into two: primary and secondary projects.

### 2.3.1 Asteroid Search (primary)

Description: This project is run by the International Astronomical Search Collaboration (IASC) and PACS e-Lab is their biggest partner in Africa. The project is also a prerequisite for all projects. In this area, individuals forming a group are trained on how to use the Astrometrica program to search for asteroids with practice datasets. They learn how to analyze these datasets, prepare a Minor Planet Center Report, and submit it back to IASC. PACS e-Lab then proceeds to register these groups on IASC's Pan African Asteroid Search Campaign group to receive fresh data to analyze. The campaign is monthly, and each team receives a certificate of participation from IASC. A week later, their reports are evaluated by IASC for true or false asteroid reports. If they are true, they become preliminary discoveries. These will be further evaluated for six months to one year to determine if they are real asteroids. If they are, they become provisional discoveries and will be given provisional numbers and cataloged. After some years, the team will be able to name their discoveries. Additionally, the asteroid research citizen science project is part of NASA's planetary defense program, which aims to monitor asteroids that can potentially hit Earth in the future. Further details about this project are highlighted in another paper titled "*Pan-African Asteroid Search Campaign: Africans' Sheer Contribution to Planetary Defense*".
Training methodology: Recorded videos and online training sessions.
Impact: Since the inception of PACS e-Lab on December $4^{th}$, 2020, over 30 provisional asteroids have been discovered by the participants.

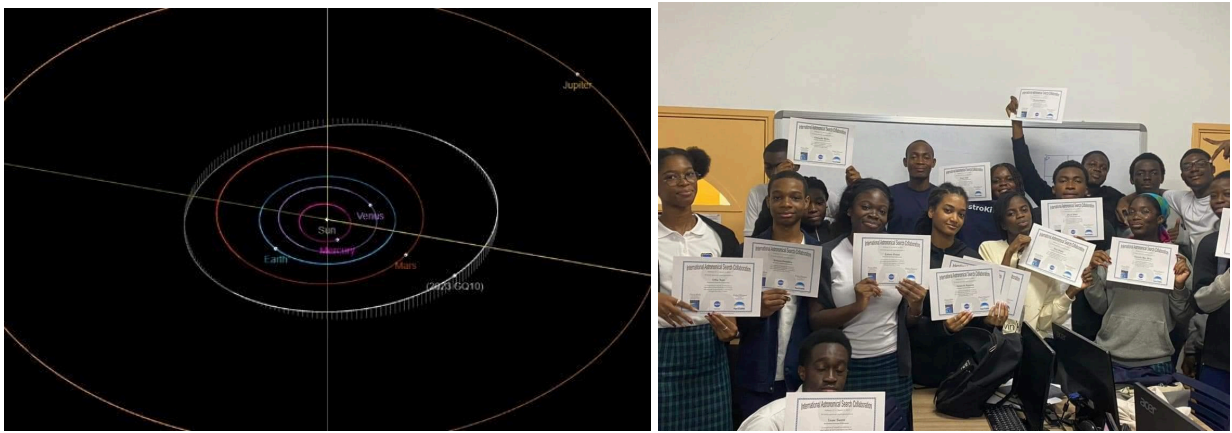

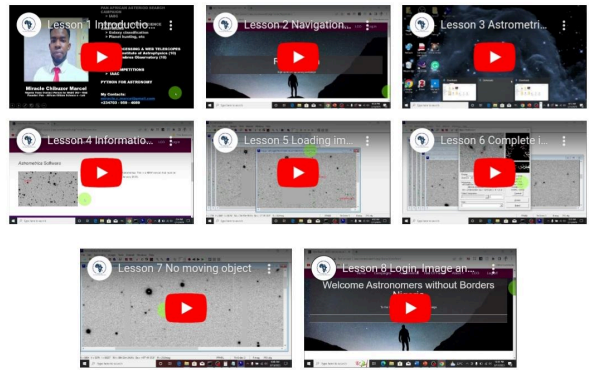

*Figure 3: The first picture is an aerial map of one of the many asteroids discovered by one of PACS e-Lab teams, and the second picture is students posing with their IASC certificates of achievement after their research. The third is a screenshot taken from the PACS e-Lab website showing the recorded practice tutorials for the asteroid search*

2.3.2 Exoplanet & Photometry (primary)

Description: This project is conducted by NASA Exoplanet Watch (Zellem, 2023), and PACS e-Lab is contributing by involving African citizen scientists. The initiative serves as a follow-up study to previously identified planets by the TESS and Kepler missions. Its primary objective is to contribute to the detection of exoplanets and refine the Mid Transit Time for these celestial bodies. This refinement is crucial for upcoming missions involving the James Webb Space Telescope and the Nancy Grace Roman Space Telescope (Zellem et al., 2020). Participants receive training on observing and capturing images of stars using the MicroObservatory (MOBS) web telescopes (Sadler et al., 2001; Gould et al., 2006; Dussault et al., 2018), focusing on stars with potential exoplanets. The MicroObservatory (MOBS) is owned and operated as a free educational service by the Center for Astrophysics | Harvard & Smithsonian. Additionally, they learn to conduct photometry using the EXOplanet Transit Interpretation Code (EXOTIC) program, a Python 3-based data reduction pipeline that is mounted on Google Colab designed for citizen scientists (Zellem et al., 2020) to be used to generate exoplanet light curves. The acquired skills also include submitting observation reports to the American Association of Variable Star Observers (Olcott, 1942). Importantly, this activity is not time-limited and can be undertaken whenever the team has available free time. Further details about this project will be highlighted in another paper. Training methodology: Recorded videos but with plans for online training sessions and workshops.

Impact: This project was incorporated in January 2024, therefore it has just begun. Only a handful of the participants have been able to observe and reduce exoplanetary datasets. But PACS e-Lab has plans to schedule live training sessions and workshops to disseminate this project.

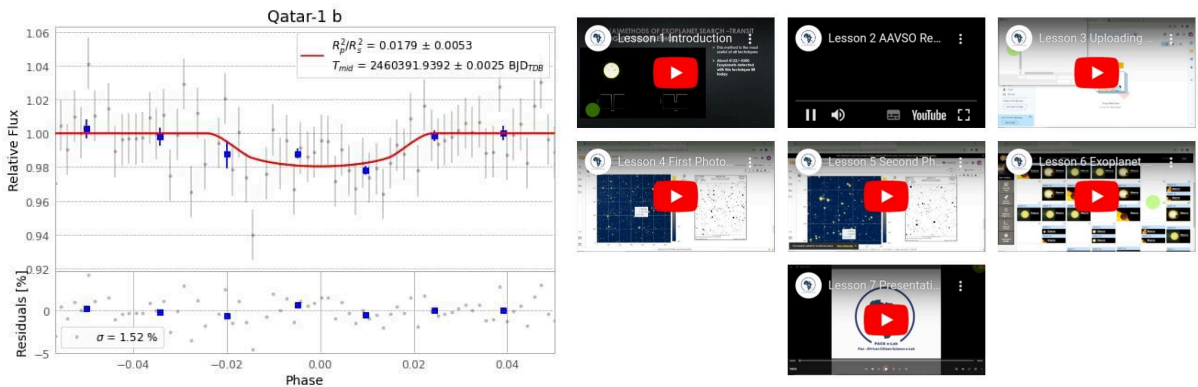

*Figure 4: One of the light curves from a transiting exoplanet produced by one of the citizen astronomers. The second picture is a screenshot taken from the PACS e-Lab website showing the recorded practice tutorials for exoplanetary research.*

2.3.3 Astronomy Research writing (primary)

Description: This project is meant to expose African citizen astronomers to astronomy research that would lead to an actual peer-review publication. PACS e-Lab has been able to curate a few soft astronomy research topics that participants can study which do not require a professional level of astronomy knowledge. One such topic is in double star research, which involves the observation of double stars to update their position angle and separation in the Washington double star catalog. Citizen researchers use the Las Cumbres Observatory (LCO) for data acquisition and AstroimageJ to take measurements. Further details about this project will be highlighted in another paper.
Training methodology: Recorded videos and online interaction with the research supervisor.
Impact: At the time of composing this literature, 10 manuscripts have been submitted with a few published in the Journal of Double Star Observation.

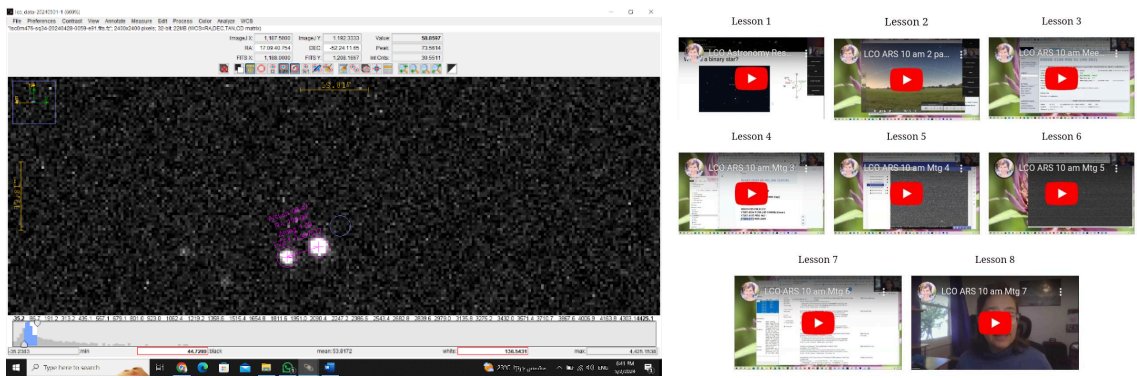

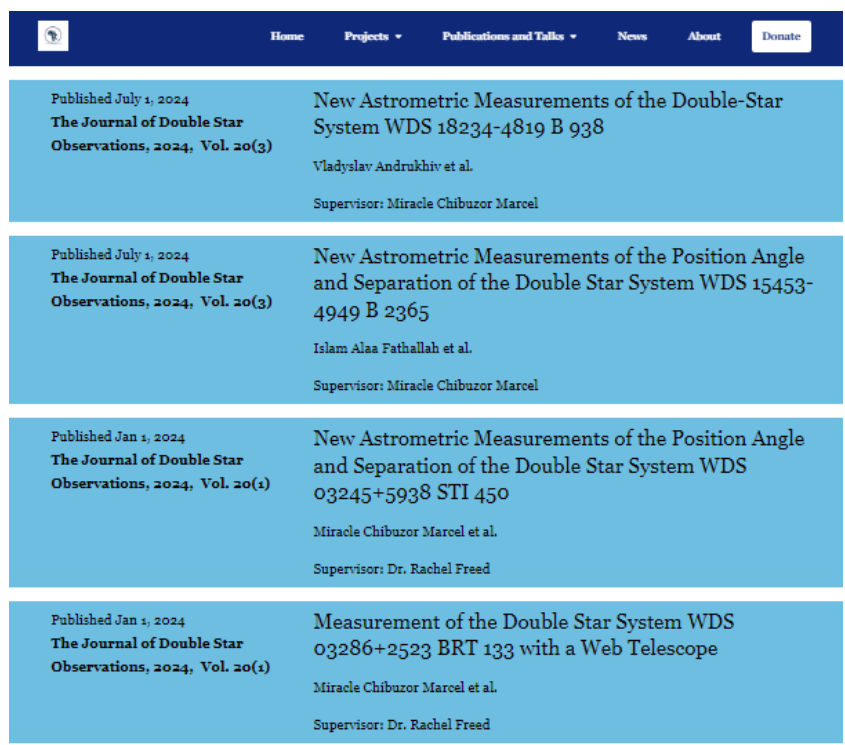

*Figure 5: The first image shows the process of Astrometry on double stars with AIJ by one of the citizen researchers. The second is a screenshot taken from the PACS e-Lab website showing the recorded videos of the double-star research procedure by Dr. Rachel Freed. The third is a screenshot from the PACS e-Lab website of some of the student-led publications.*

2.3.4 Astro Photo Visual Development (primary)

Description: Unlike the other research-based projects, this is not a research activity, but an important skill for amateur astronomers to process deep space images to produce beautiful visuals. The participants are trained on how to retrieve archival datasets from the Hubble Space Telescope (Scoville et al., 2007), James Webb Space Telescope (Gardner et al., 2006), and Las Cumbres Observatory (Brown et al., 2013). They learn how to acquire these datasets and process them with software like Photoshop, Fits Liberator, Siril, Gimp, etc. to produce colorful visuals. Further details about this project will be highlighted in another paper
Training methodology: Recorded videos but with plans for online training sessions and workshops.
Impact: This project was developed and incorporated in May 2024, therefore it has just begun the period of writing this literature. Only a handful of participants have been able to retrieve/observe and process the images. But PACS e-Lab has plans to schedule online training sessions and workshops to disseminate this project.

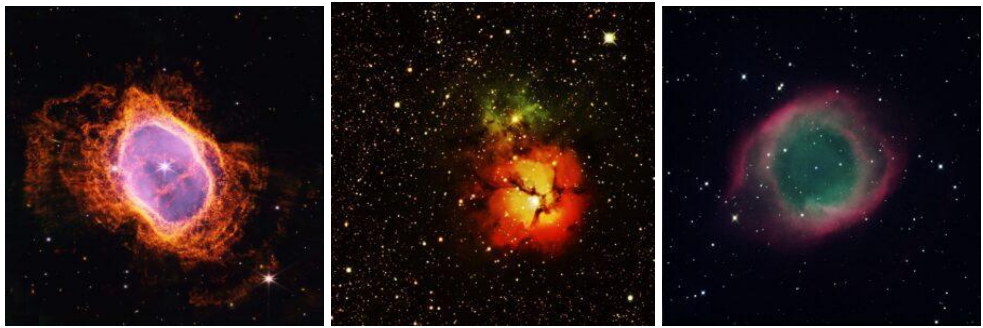

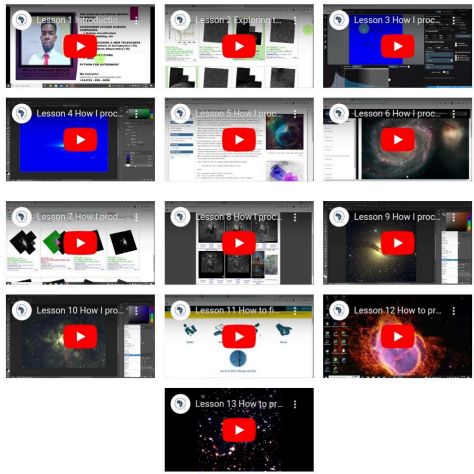

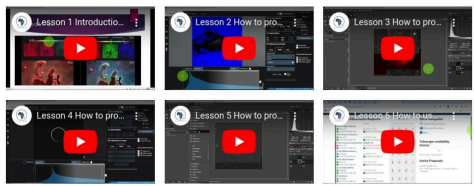

*Figure 6: Space images from the JWST & LCO processed by the citizen astronomers. The second is a screenshot taken from the PACS e-Lab website showing the recorded tutorials both for Photoshop and Gimp on Astro Photo Visual Development skills.*

2.3.5: PACS e-Lab Give-Away (secondary)

Description: PACS e-Lab leverages the support of its regional and global networks to support team members through donating telescopes. Some of PACS e-Lab's collaborators include the Jean Pierre Grootaerd Telescope for all programs and the African Astronomical Society. This is PACS e-Lab's effort to encourage its citizen Astronomers to be active in astronomy outreach in their localities as they continue to participate in the entity's advanced projects. The focus is mostly on countries where there were struggles recruiting individuals to participate in the projects because there are no basic astronomy awareness campaigns ongoing there. These telescopes when used for public outreach will spark curiosity and interest in astronomy and space science in those countries. Further details about this project will be highlighted in another paper

Impact: PACS-Lab has donated five telescopes to its citizen scientists in Malawi, Botswana, Zimbabwe, Uganda, and Lesotho. PACS e-Lab is committed to continuing in these steps until all deserving and motivated teams get all the tools they need for outreach.

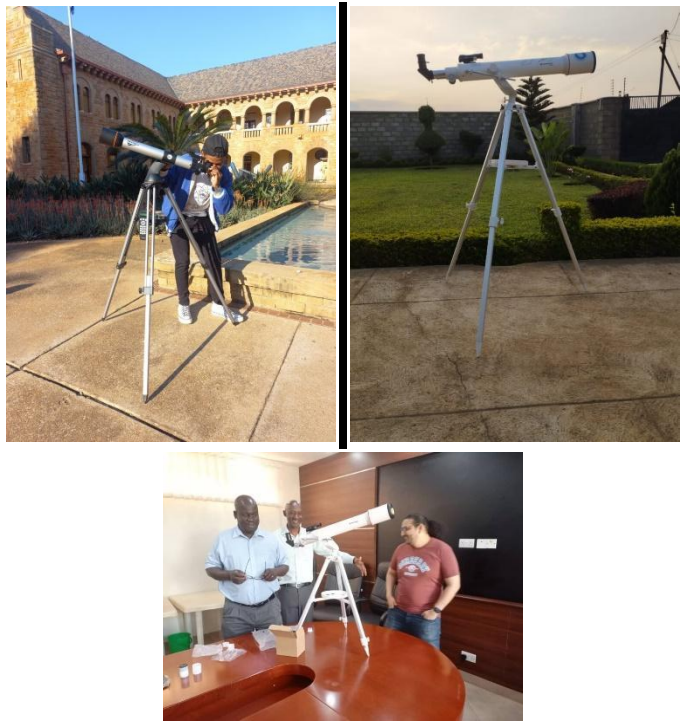

*Figure 7: Telescope donations to our Lesotho, Malawi, and Uganda teams*

2.3.6 ARISS Event (primary)

The Amateur Radio on the International Space Station (ARISS) program is an initiative that allows educators, students, space enthusiates to communicate directly with astronauts aboard the International Space Station (ISS) via amateur radio (Bauer, 2019). This is PACS e-Lab's latest project that will allow African teachers, students, and space enthusiasts the opportunity to interact with astronauts aboard the international space station with amateur radio just like students from developed countries. PACS e-Lab's application was recently accepted and plans are being made for this event which is set to be between January – June 2025. However, PACS e-Lab is forging ahead to make it a recurring activity on the African continent. Further details about this project will be highlighted in another publication.

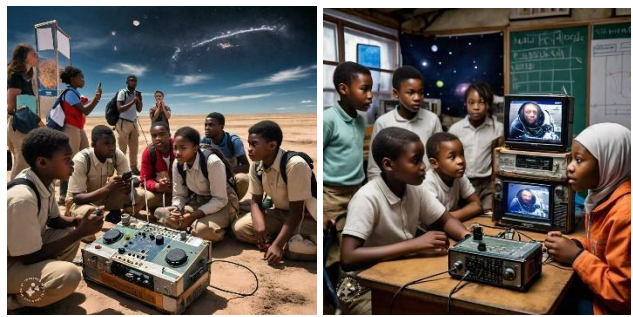

*Figure 8: AI-generated images of how the ARISS event will look like*

## 2.4 Access to Astronomical Data

Among all research endeavors, astronomy is among those that provide huge amounts of data. Spanning from both orbiting space telescopes and ground-based telescopes. Most of these observational data are freely available to the public. PACS e-Lab utilizes some of these for its endeavors.

Asteroid Search: For our asteroid search endeavors, data is supplied from the Panoramic Survey Telescope & Rapid Response System (Pan-STARRS) (Chambers et al., 2016) at the Institute for Astronomy at the University of Hawaii and the Catalina Sky Survey (CSS) (Larson et al., 2003) at the Lunar & Planetary Laboratory at the University of Arizona to the International Astronomical Search Collaboration (IASC). IASC then distributes the data to their registered recruiting groups including PACS e-Lab. After measurement, the report called the Minor Planet Center (MPC) report, is submitted back to IASC.

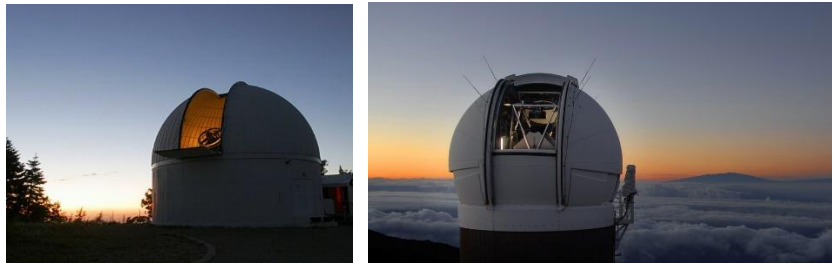

*Figure 9: Pan-STARRS right and CSS left*

Exoplanet & Photometry: In this project, exoplanet datasets are supplied from three sources: NASA Observation Data Checkout, which works by submitting an email address through the NASA Exoplanet Watch website; the 0.15m MicroObservatory managed by the Center for Astrophysics | Harvard & Smithsonian for observation of planetary transits; and the 0.4m Las Cumbres Observatory (Brown et al., 2013) for observation of planetary transits too.

Astronomy Research Writing: In this project, observations are made using the 0.4m LCO only. It should be noted that PACS e-Lab recently became one of the Las Cumbres Observatory Global Sky Partners. This opportunity allows the team to utilize the 0.4m telescope for observations.

Astro Photo Visual Development: In this project, PACS e-Lab utilizes archival data from the Hubble Space Telescope and James Webb Space Telescope. PACS e-Lab teams can also schedule astrophotography missions using the 0.4m LCO. The entity also has access to the Slooh telescopes for live observation of astronomical events like eclipses.

**2024 North American Solar Eclipse**

Use the links to view and download all images we captured with our observation time on the Slooh telescopes. The below images are just samples. Before Totality, During Totality, and After Totality.

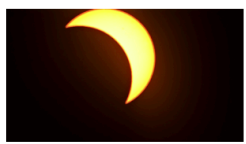
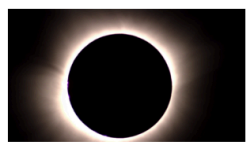
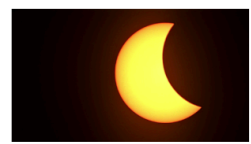

Before Totality | During Totality | After Totality

Figure 10. *A screenshot which was taken from the PACS e-Lab website showing the April 8, 2024, North American eclipse images captured by the PACS e-Lab team using the Slooh web telescopes.*

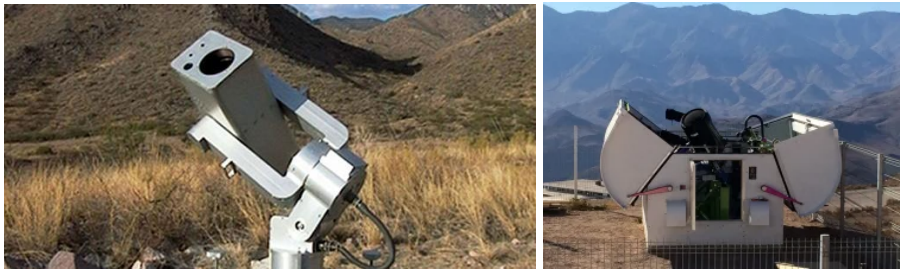

*Figure 11: One of the networks of the 0.15m MOBS left and the 0.4m LCO right*

## 2.5 Determining impacts

To determine the impact of PACS e-Lab endeavors over these years, the team leadership of each recruiting unit across the continent was, for the first time since the inception of PACS e-Lab, issued a survey. They were required to fill it out by answering these questions: 1. The full name(s) of the team leadership. 2. The names of their countries and geographical zones. 3. The first year the group joined PACS e-Lab. 4. Name of the group/school/club/organization. 5. Age distribution of their participants. 6. Full names, gender identity, highest level of education, year of entry, and projects participated in, of all team members of that unit. 7. Comments on the experiences of team members towards the projects. 8. How the unit is adopting PACS e-Lab projects to the classroom, club, or community. 9. If the team has participated in any other projects outside PACS e-Lab. 10. Some of their challenges in participating or recruiting participants. 11. Their future plans regarding their participation and dissemination of the projects.

Over 70 responses were received after a week, and these formed the data through which PACS e-Lab impacts can be interpreted. These are shown in the results section below.

## 3.0 Results

Table 1 shows the summary of the survey responses from the teams' leadership. Table 2 shows the African countries connected to PACS e-Lab, classified under their geographical zones. The names of these units are listed alongside the unit leadership. The data from the survey were also used to prepare the charts below. Figure 18 is the picture of the African map showing PACS e-Lab's citizen science networks.

*Figure 12: The screenshot of the survey is summarized in the Google sheet.*

Table 1: Shows African Countries connected to PACS e-Lab classified under their geographical zones. The names of these units are listed alongside the unit leadership.

| Zones | Country | Unit Names | Unit leadership |
|---|---|---|---|
| North Africa | Algeria | Algeria AstroResearchers (C) | Amina Ramadan |
| | Egypt | IEEE- A student breach activity (B) | Mohamed Elattar |
| | Egypt | Pharaohs of Space (C) | Kareem Waleed Hosny |
| | Egypt | Space -Water-Environment Nexus e - Center (B) | Dr. Sherine Ahmed El Baradei |
| | Libya | Amateur Astronomy Libya (C) | Saja Alhoush Sulayman |
| | Morocco | Steps Into Space Team (SIS) (C) | Salma Regaibi |
| | Morocco | Oukaimeden Observatory (A) | Dr. Meryem Guennoun |
| | Tunisia | Tunisian Astronomical Society (A) | Nejmeddine Etteyeb |
| | Tunisia | Aljarid Astronomie (C) | Ben Abdallah ridha |
| West Africa | Benin | Sirius Astro-Club Benin (C) | Mawugnon Axel Gahou |
| | Benin | CosmoLAB Hub Association (C) | Thierry Martial TCHANGOLE |
| | Burkina Faso | Young Burkinabe Astrophysicists (B) | BADO Fabrice |
| | Côte d'Ivoire | Association Ivoirienne d'Astronomie (B) | Dr. Kassamba Abdel Aziz Diaby |
| | Gambia | Physics Dept. University of Gambia (B) | BABOUCARR M. B BOBB & Kebab Jaiteh |
| | Ghana | PRAGSAC (A) | Matthias Asare-Darko |
| | Ghana | Ghana Planetarium (C) | Sarah Abotsi-Masters |
| | Guinea | UNESCO Center For Peace USA in Guinea (A) | Mamadou Balde |
| | Liberia | Center for Science Education (A) | Allen Thomas |
| | Niger | Niger Space Surfer (B) | Dr. Souley Yahaya |

|  | Nigeria | The Astro Group of the Rivers State University (B) | Dr. Esaenwi Sudum |
|---|---|---|---|
|  | Nigeria | Kano Space Hub (C) | Sulaiman Muhammad Gumau |
|  | Nigeria | Astronomers Without Borders Nigeria/PACS e-Lab Nigeria. (C) | Dr. Olayinka Fagbemiro |
|  | Nigeria | Asteroid Research Group, Chukwuemeka Odumegwu Ojukwu University Uli Campus Anambra State (B) | Ngozika Frances Anekwe |
|  | Nigeria | AbbaGod's Space (C) | Chidera Emmanuel Abachukwu |
|  | Nigeria | Spiritan University Nneochi (B) | Dr. Chinedu Nwadike |
|  | Senegal | Orion Astrolab and Department of Physics, University Cheikh Anta Diop (B) | Dr. Salma Sylla |
|  | Sierra Leone | Cosmic Gazers Research Institute Sierra Leone (B) | Mohamed Ali Koroma |
|  | Sierra Leone | Sierra Leone Geospatial and Space Agency (A) | Dr. Reginald Hughes & Mamaja Jalloh |
|  | Togo | NGO Science Géologique pour un Développement Durable (SG2D) (C) | Christian Kontoa Koussouwa Gbaba |
| Central Africa | Angola | Angolan Space Program Management Office (GGPEN) (A) | Dr. Gilberto Gomes |
|  | Angola | Associação Angolana de Astronomia (C) | Moisés da Silva |
|  | Chad | Toumaï (C) | Mamadou Mahamat Djabbi |
|  | Cameroon | Astronomy Club of the University of Buea (B) | Tengwi Mogou Ornela |
|  | CAR | IT Dreams and Promises of the University of Bamgui (B) | MAMOUDOU KOUROUMA |
|  | Congo | University Marien Ngouabi (B) | Jocelyn Franck-Patient and DINGA Jean Bienvenu |
|  | Congo | Denis Sassou N'GUESSO University (B) | Jocelyn Franck-Patient and DINGA Jean Bienvenu |
|  | DRC | Astroclub Kongo Central (C) | MAFUKA LUSALA JOSEPH |
|  | DRC | Lubumbashi Astro Club (C) | Dieumerci kaseha |
|  | Gabon | NAEC - Gabon (B) | Dr. Hans ESSONE OBAME |
| East Africa | Burundi | Physics Dept. University of Burundi (B) | Dr. René Simbizi & Dr. Salomon Mugisha |
|  | Ethiopia | Ethiopia Space Science Society Citizen Science Team (A) | Kirubel Memberu & Lidia Dinsa |
|  | Kenya | Space Partnerships and Applications Company Kenya (C) | Dennis Kiyeng |

|  | Kenya | Kenya Space Agency - Education and Outreach Students' Network (A) | Harold Safary |
| --- | --- | --- | --- |
|  | Kenya | Amateur Astronomical Society of Kenya (C) | Samuel Nyangi |
|  | Mauritius | IAU - NOC Mauritius (A) | Arvind Ramessur |
|  | Madagascar | Haikintana Astronomy Association (C) | Andoniaina Rajaonarivelo |
|  | Rwanda | Physics Department, University of Rwanda, College of Science and Technology (B) | Dr. Ntahompagaze Joseph |
|  | Sudan | Sudanese Asteroid Hunters (C) | Ahmed Saeed |
|  | Sudan | Institute of Space Research and Aerospace (ISRA) (A) | Mohammed Yahya Alradi Eldaw |
|  | Somalia | Somali Space Explorers (C) | Ictisam abdulkadir mohamoud |
|  | South Sudan | Mayardit Academy for Space Sciences - University of Juba (A) | Dr. Christopher Tombe Louis Gore |
|  | Seychelles | Citizen Science Seychelles (C) | Erias Kasule |
|  | Tanzania | Department of Physics, University of Dodoma (B) | Privatus Pius |
|  | Uganda | National Curriculum Development Centre of Uganda (A) | Betty Rose Nabifo |
| Southern Africa | Zambia | Students from the Copperbelt University (B) | Tshombe Nkhowani & Lhozindaba Zulu |
|  | Zambia | Zambia Space Explorers (C) | Henry Sanderson Viyuyi |
|  | Zambia | Kabulonga Girls' Secondary School (C) | Tresford Chilufya Kalunga |
|  | Mozambique | Detetives do Cosmos (C) | Nunes Alfredo Assano and Edson Domingos Jequecene |
|  | Malawi | Celestial Explorers (C) | Bauleni BVUMBWE |
|  | Zimbabwe | Zimbabwe National Geospatial and Space Agency (A) | Tawanda Kelvin Simon/ Fortune Tatenda Matambo, |
|  | Zimbabwe | University of Zimbabwe (B) | Tawanda Kelvin Simon/ Fortune Tatenda Matambo |
|  | Zimbabwe | Zimbabwe Astronomical Society (A) | Thobekile Sandra Ngwanw |
|  | Botswana | Marang Junior Secondary School (C) | Nonofo Angela |
|  | Botswana | University of Botswana Astronomy Club (B) | Molly Nkamogelang Kgobathe |
|  | Namibia | Department of Physics, Chemistry and Material Science University of Namibia (B) | Immanuel Luis & Sigrid Shilunga |
|  | South Africa | BlueCraneSpace Astronomy & Astrophysics Department - University of Pretoria (B) | Keketso Qhomane |
|  | South Africa | Night Sky Tours (C) | Oosthuizen Leonardo |

|   | South Africa | Galaxy Explorers (C) | Bongiwe Portia Mkhabela |
|---|---|---|---|
|   | Lesotho | Maloti Space Explorers (C) | Keketso Qhomane |
|   | Eswatini | Destiny Stars of the University of Eswatini (B) | Dr. Mdumiseni Wisdom Dabulizwe Dlamini |

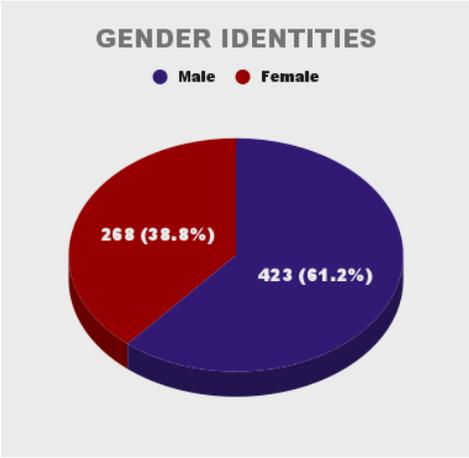

*Figure 13: Pie chart displays the Gender identity of our African Citizen Astronomers*

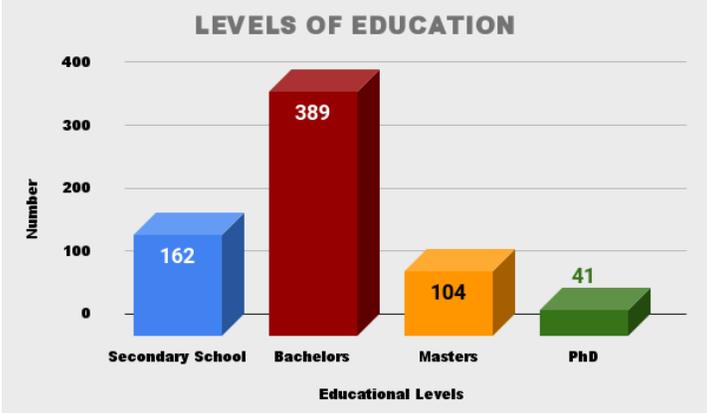

*Figure 14: Chart showing the level of education of the citizen Astronomers*

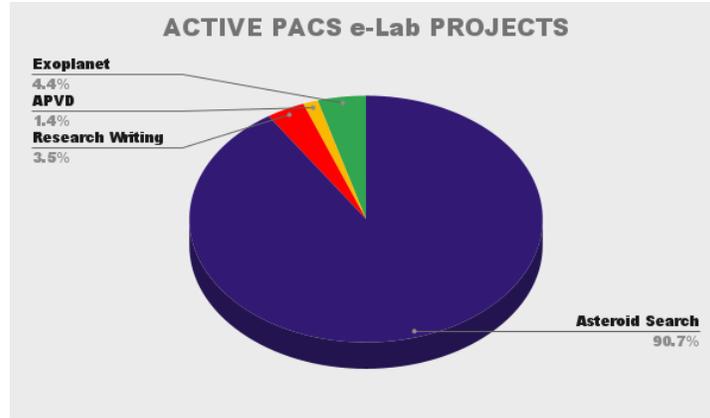

*Figure 15: Chart of the active PACS e-Lab projects participation by statistics*

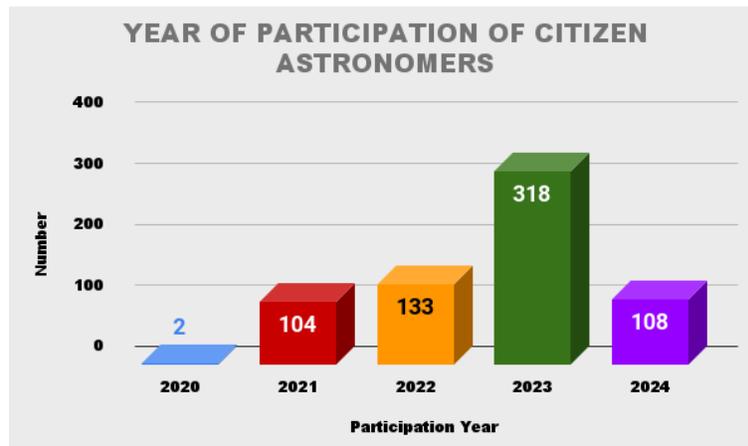

*Figure 16: Chart showing the number of citizen astronomers recruitment each year.*

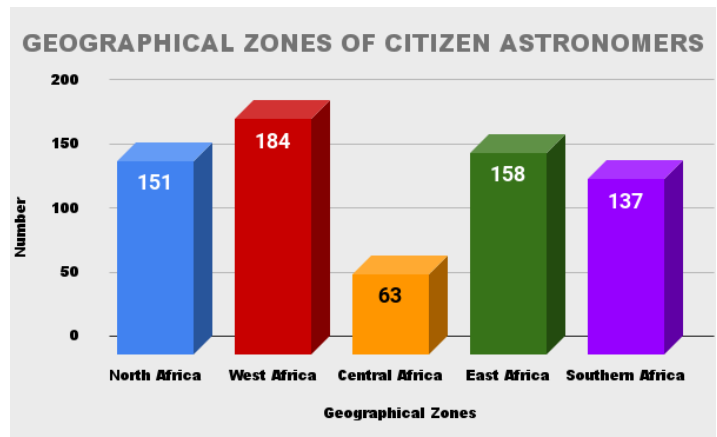

*Figure 17: Shows the number of recruited citizen astronomers per geographical zone.*

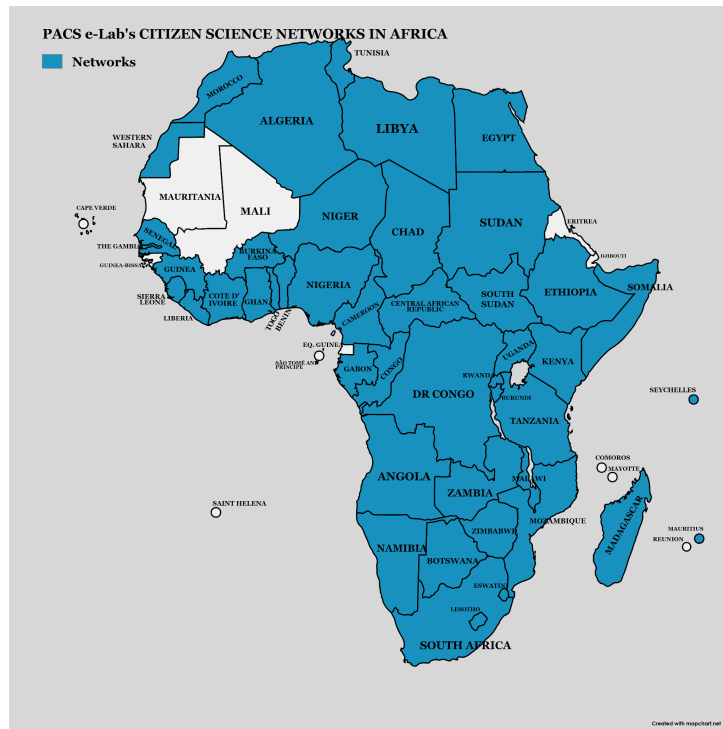

*Figure 18: African map showing PACS e-Lab's citizen science networks*

## 4.0 Discussions

The survey identified 70 units or groups from 45 African countries. The total number of citizen scientists recorded is 691. In Table 2 above, each of the units were classified as A, B, and C to make them easier for readers to distinguish. A = 15 corresponds to entities such as national space agencies, national astronomical societies, national observatories, and other national entities or representations; B = 25 represents university groups; and C = 30 units are secondary schools and private organizations.

Figure 13 shows the gender identities of the participants, indicating that they identified as either male or female, however, PACS e-Lab has made emphasis on welcoming all persons irrespective of their gender identities. Overall, the number of male participants surpasses that of female participants. Moreover, Table 1 shows that some groups have a higher number of women and girls participating, while others have a close margin, and most have a higher male-to-female ratio.

Figure 14 shows the educational levels of the citizen scientists, with those holding first degrees leading, followed closely by secondary school students and those with master's degrees. The PhD holders are mostly tertiary-level teachers who have adopted the projects into their classrooms to enhance their teaching experiences.

Figure 15 shows that the asteroid search is the leading project, followed by others like exoplanet research, research writing, and astrophotography. The reason for this is that the asteroid search was the first and foremost project of PACS e-Lab over the years. The rest were recently developed and incorporated. Another reason is that, according to PACS e-Lab management, the asteroid search is a prerequisite for all the projects and should be undertaken before introducing other projects to any participant. This will better prepare them for other projects. However, this rule is not strict.

Figure 16 shows the entry year of citizen scientists starting from the inception of PACS e-Lab in 2020 to June 2024, the time of writing this literature. It shows a steady increase in the recruitment and training of new citizen astronomers. However, the sharp drop in 2024 is due to two reasons: in late 2023 and early 2024, PACS e-Lab management spent most of the time developing and incorporating new projects, such as astrophotography (also called Astro Photo Visual Development in PACS e-Lab), exoplanet and photometry, double star research, ARISS event, and a host of other projects still in the developmental pipeline. Another reason is that the time of writing this literature is June, which is still the middle of the year, so it is too early to pass judgment on the low number of recruits. In the second quarter of 2024, PACS e-Lab went into a partnership with the International Astronomy and Astrophysics Competition (IAAC), which has one of the largest numbers of Africans participating in their astronomy competitions, to use their platform to recruit their African participants into PACS e-Lab. This effort alone is expected to boost the overall recruitment of 2024.

Figure 17 shows the total number of recruited citizen scientists in the five geographical zones of Africa: North Africa, West Africa, Central Africa, East Africa, and Southern Africa. West Africa took the lead, followed by East Africa, NorthAfrica, Southern Africa and lastly Central Africa. No specific reason could be pointed out as to why this is so because the PACS e-Lab management sends invitations equally to everyone across the continent. There are also almost equal limitations in terms of access to the internet, electricity, and language barriers, which the management applies almost the same solution to resolve.

Figure 18 shows the map of Africa. The 45 countries in blue are those with connections to PACS e-Lab, while the rest are not yet. The PACS e-Lab management continues to seek more collaborations in those countries not yet under their radar.

### 4.1 The Other Survey Responses

### 4.1.1 Age Distribution

The age distribution stated in the survey includes children, defined as between 3-12 years; teenagers, defined as between 13-19 years; and adults, defined as above 20 years. Out of all 70 teams, 3 teams identified as having a mixture of children, teenagers, and adults; 24 teams have a mixture of teenagers and adults; 42 teams are primarily composed of adults; and one group is entirely made up of teenagers.

### 4.1.2 Participation in other astronomy and space science projects outside PACS e-Lab's

The teams were asked to state if they had ever participated in any other astronomy or space science hands-on projects outside of the PACS e-Lab, but the majority stated they had not and PACS e-Lab's projects were their first experience.

### 4.1.3 Comments on the experiences of team members from different units towards the projects.

Mohamed Moustafa Elattar affiliated with the IEEE - A student breach activity in Egypt reported that most participants were very interested in the experience, particularly in asteroid discovery. They developed skills such as teamwork, time management, searchability, and attention to detail.

Dr. Ntahompagaze Joseph, affiliated with the Physics Department at the University of Rwanda, College of Science and Technology, mentioned that the team has found great enjoyment in participating in asteroid searches across various rounds since 2022 they joined PACS e-Lab. They have developed skills in using Astrometrica software and have honed their abilities in planning image processing and reporting.

Mrs. Joy Olayiwola, affiliated with the Nigeria Space Agency (NASRDA), says she has had an exhilarating journey so far. Although she did not study astronomy or its core subjects at the undergraduate level, joining PACS e-Lab has boosted her love for astronomy. She gets excited at the mention of the name and is proud to be a citizen scientist in astronomy-related projects.

Ms. Mary Ekwu from Nigeria says her experience has been one of great learning and many expository sessions. Space science or astronomy seemed abstract and far-fetched until she came in contact with PACS e-Lab. It has been a dream come true, and she now loves space science even more.

Elijah Ibharalu from Nigeria says that PACS e-Lab has been of immense benefit to him. He now has ideas and can work on projects in space research that he never thought he would be able to do, with just a little bit of hard work.

Keketso Qhomane, an undergraduate student from the Blue Crane Space Astronomy & Astrophysics Department at the University of Pretoria, and team leader of the project, stated that the Asteroid Search project sparked a profound interest in space sciences, transforming ordinary society members into citizen astronomers involved in planetary defense. The team members hail from three different Southern African countries: South Africa, Lesotho, and Zimbabwe. This project has significantly elevated the society's perception of citizen science and astronomy. As the first team leader, Qhomane observed the next generation of leaders continuing his work since 2023, with the team making two preliminary asteroid discoveries in three search campaigns that year, and adding two more in four campaigns in 2024. Qhomane expressed pleasure in seeing his society mates following the path he helped forge since Blue Crane Space began participating in the IASC Asteroid Search Campaigns in April 2023.

Tawanda Kelvin Simon, affiliated with the Zimbabwe National Geospatial and Space Agency, stated that his team has developed an increased interest in the space world, particularly the astronomy industry.

Dr. Sherine Ahmed El Baradei, affiliated with the Space -Water-Environment Nexus e - Center, Egypt, stated that the asteroid search project has significantly benefited her team by opening new horizons for understanding astronomy.

Bongiwe Portia Mkhabela from South Africa said it was a fun experience. She enjoyed working with her peers and learning how to use Astrometrica.

Salma Regaibi, the president of the Steps Into Space Team (SIS) in Morocco, said, "Since we are so passionate about Astronomy, it was an amazing experience during which we learned many scientific things, especially how to use the software for asteroid research using NASA images."

Christian Kontoa Koussouwa Gbaba, affiliated with the NGO Science Géologique pour un Développement Durable (SG2D) in Togo, says, "They all enjoyed learning to use the Astronomica software and the entire asteroid research process. Our campaign resulted in the discovery of two provisionally classified asteroids, which further increased the members' interest in astronomy and space sciences."

Dr. Meryem Guennoun, an astrophysicist affiliated with the Oukaimeden Observatory in Morocco, says, "My team members were already very passionate about astronomy. I was supervising them in their school's STEM club. They were very motivated to be part of the project, learn new things, actually practice astronomy, and get a certificate for it."

Matthias Asare-Darko, affiliated with PRAGSAC in Ghana, reported that new skills were learned and team members also experienced boosts in confidence levels.

Samuel Nyangi, associated with the Amateur Astronomical Society of Kenya, highlighted that PACS e-Lab projects have significantly increased undergraduate students' exposure to scientific astronomy research and sparked a greater interest in the field of astronomy.

Dieumerci Kaseha, affiliated with the Lubumbashi Astro Club in the Democratic Republic of Congo, reported that their experience has sparked a spirit of research among participants. Additionally, they found the experience to be enriching for their team.

Andoniaina Rajaonarivelo, president of the Haikintana Astronomy Association in Madagascar, reported that through participation in PACS e-Lab projects, their team's perception of scientific research has transformed significantly. Previously, some members viewed research as a complex and distant field conducted by experts in isolated labs. However, involvement in citizen science projects has shown them the accessibility and collaborative nature of research. They now actively contribute to real scientific endeavors, feeling a sense of ownership and purpose in the analysis process.

Thobekile Sandra Ngwanw, president of the Zimbabwe Astronomical Society, reported that their team's involvement in the asteroid search project has significantly enhanced their perception of scientific research. Engaging with automated observation technologies and data analysis has provided hands-on experience, simplifying many aspects of the research process and reducing intimidation. Several participants have been inspired to pursue their research, with two members now pursuing masters in astronomy.

Participation in the International Astronomical Search Collaboration (IASC) has been particularly inspiring, as the high-quality data provided has enabled original discoveries and hands-on astronomy engagement. This direct involvement has turned abstract concepts into tangible experiences, fueling collective enthusiasm for ongoing and future space exploration initiatives. Contributing to significant scientific discoveries has built confidence and strengthened connections within the global scientific community, highlighting the collaborative nature of astronomical research and empowering team members in their pursuit of further studies and careers in the field.

Henry Sanderson Viyuyi, founder of the Zambia Space Explorers, expressed that the experience has been wonderful and life-changing for their team. The simplicity of the projects has made it easy for members to connect and understand, allowing participation regardless of their level of astronomy knowledge.

Ms. Sigrid Shilunga, a master's student in the Department of Physics, Chemistry, and Material Science at the University of Namibia and team leader, reported that the IASC Asteroid Search project has had a profound impact on their team. It has enhanced their research abilities and academic knowledge while sparking a strong interest in space exploration. The project provided practical context for their studies, demonstrating how academic disciplines apply in the real world, thereby influencing career aspirations and fostering enthusiasm for scientific research. Participation in the Asteroid Search projects also offered valuable hands-on experience, improving their understanding of scientific research processes and methodologies. Overall, the opportunity to engage in real-world space research significantly heightened their team's interest in space exploration.

Ms. Betty Rose Nabifo, affiliated with the National Curriculum Development Centre in Uganda and team leader of her team, reported that the asteroid search project has sparked significant interest among learners. There is a strong eagerness among them to learn about the composition of space.

Baboucarr M. B Bobb, an undergraduate student from the Physics department of the University of Gambia and team leader of his team, reported that the project has been an eye-opener for them as Physics students, allowing them to discover beyond the scope of the classroom.

Nunes Alfredo Assano and Edson Domingos Jequecene, affiliated with Detetives do Cosmos in Mozambique, reported that participating in these projects has reignited their team's curiosity and enthusiasm for space. They have noticed a growing interest in space exploration among participants, with some now considering careers in this field. The opportunity to contribute to real discoveries has made space feel more tangible and exciting for them.

Kirubel Memberu & Lidia Dinsa affiliated with the Ethiopia Space Science Society Citizen Science Team, noted that these campaigns have served as a platform for their team to collaborate and learn together. The shared experiences and challenges have enhanced their teamwork and communication skills. Moreover, these projects have clarified the research process, demonstrating that individuals from diverse backgrounds can make meaningful contributions to science.

Mr. Erias Kasule, a physics and mathematics teacher from Seychelles, says that the PACS e-Lab projects he has worked on so far have been an inspiration and an eye-opener to what we have in outer space. They have encouraged him to take on other projects in the same field. "Amazing and informative" is all I could say.

### 4.1.4 How the units are adopting PACS e-Lab projects to the classroom, club, or community.

Sarah Abotsi-Masters from Ghana says she regularly promotes PACS e-Lab's astronomy projects, particularly the asteroid search, to various astronomy-related WhatsApp groups she is part of. She extends these promotions to teachers, the general public, and members of their Planetarium mailing list.

Mohamed Moustafa Elattar from Egypt writes that he regularly promotes PACS e-Lab's astronomy projects by posting about them on social media and verbally encouraging others to participate. He emphasizes the skills, advantages, and enjoyment that come from taking part, especially for those interested in astronomy.

Dr. Ntahompagaze Joseph, affiliated with the Physics Department at the University of Rwanda, College of Science and Technology, mentioned that he includes asteroid search activities in the curriculum of the Astronomy and Astrophysics course for third-year Bachelor of Physics students.

Privatus Pius, affiliated with the Department of Physics at the University of Dodoma, Tanzania, has initiated special lectures in his regular classroom to explain PACS e-Lab's projects. This effort aims to raise awareness and encourage student participation, resulting in an increased number of participants.

Dr. Esaenwi Sudum, affiliated with the Department of Physics at Rivers State University, Nigeria, mentioned that their Astro Group has successfully integrated PACS e-Lab programs into the department's teaching curriculum for both MSc and BSc students in Astrophysics. This initiative has made it compulsory for Astronomy students to engage in basic research through these programs.

Dr. Souley Yahaya, affiliated with the Niger Space Surfer, says his team organizes periodic outreach events and special sessions to explain their activities. These events aim to inform the public about their initiatives and invite astronomy enthusiasts to join their efforts.

Ben Abdallah Ridha, affiliated with Aljarid Astronomie in Tunisia, says that they define PACS e-Lab's objectives and then present space and asteroids, including their concept and the method of monitoring these asteroids through a digital system to the participants.

Dr. René Simbizi, affiliated with the Physics Department at the University of Burundi, mentions that he is adopting PACS e-Labs projects in the classroom while teaching physics and astronomy.

Ngozika Frances Anekwe, affiliated with the Physics Department, Chukwuemeka Odumegwu Ojukwu University, Uli Campus, Anambra State, Nigeria, mentions that PACS e-Labs projects have helped explain astronomy in the classroom through hands-on practical.

Tawanda Kelvin Simon, affiliated with the Zimbabwe National Geospatial and Space Agency, stated that his team tries to include most of the information from the Asteroid Search project in all their outreach programs aimed at enlightening people about space.

Dr. Sherine Ahmed El Baradei, affiliated with the Space -Water-Environment Nexus e - Center , Egypt, says she will be teaching a course entitled "Astronomy and Space for Water and Environmental Engineering" in the coming fall semester. The course will include the experience of the center as well as asteroid search, and she plans to engage her students in the asteroid search campaign as a project.

Tshombe Nkhowani from Zambia says he teaches his secondary school students how to use the Astrometrica software for asteroid searches.

Dr. Meryem Guennoun says she is implementing this project in every outreach activity she conducts and in the clubs she supervises.

Andoniaina Rajaonarivelo from Madagascar says his team dedicates specific club meetings for project participation, guiding members through the project's instructions and data analysis processes.

Mohamed Ali Koroma, affiliated with the Cosmic Gazers Research Institute in Sierra Leone, says his unit is adopting PACS e-Lab projects into the classroom, club, and community through several strategies. In the classroom, they integrate project activities into the curriculum, allowing students to apply theoretical concepts in a practical context. For example, they use the projects to demonstrate principles of physics and astronomy, encouraging students to engage in hands-on experiments and observations. He also speaks to high school students about the project research and its beauty.

Thobekile Sandra Ngwanw says they are using the asteroid search campaigns to show students in their department what using astronomy software is like. This campaign gives students access to real astronomical data, which is very exciting. They hope it will encourage students to pursue a career in astronomy, despite the limited opportunities in their country.

Henry Sanderson Viyuyi from Zambia says his team conducts outreach programs where they talk about the projects and offer hands-on experiences to pupils and students.

Molly Nkamogelang Kgobathe from Botswana says her team has been participating by giving talks locally within the university community as a starting point, and now they are giving online talks nationally and internationally through the Astronomical Society of Botswana platform. These talks focus on the participation of their members in the projects they have been involved in so far.

Kirubel Memberu & Lidia Dinsa reported that the campaign is currently being run by members of the Ethiopian Space Science Society. They have actively promoted the campaign within their organization but plan to expand their efforts beyond their current membership base. Their next step involves reaching out to a broader audience outside of their organization.

Dennis Kiyeng from Kenya says his team plans to reintroduce the program in primary and high schools, starting with the schools where they have established space clubs.

Kareem Waleed Hosny, the team leader of Pharaohs in Space, says, "In Egypt, space activities are rare and hard to find, so we are working to engage the youth population in Egypt with these PACS e-Lab projects."

### 4.1.5 Their future plans regarding their participation and dissemination of the projects.

When asked about their future plans for the projects, most responses centered around continued participation and expansion to accommodate more members of the public. This includes reaching out to local government levels, involving local chiefs, and engaging at the national level with government agency participation. Some units mentioned participating in other PACS e-Lab projects, such as Exoplanet and Photometry, Astrophoto visual development, research writing, ARISS event etc. Others proposed incorporating all these PACS e-Lab projects into their class activities for bachelor's and master's students.

There were also plans to start astronomy clubs in schools that are underrepresented in STEM. Some participants talked about sending newsletters to keep their participants informed periodically. Finally, some planned to document and share the outcomes and successes of the projects through reports, publications, and presentations at conferences, thereby contributing to the broader scientific community and inspiring similar initiatives elsewhere.

### 4.1.6 Some of their challenges in participating or recruiting participants.

The teams were asked to state their challenges in participating or recruiting participants. Their responses centered around several key points. One major issue was the low interest in astronomy and space science, compounded by widespread misconceptions about the field. Additionally, many teams faced practical challenges such as a lack of electricity and strong internet services, as well as the high cost of internet access. The lack of computers was another significant barrier.

Furthermore, the online nature of the platform was often less motivating to students compared to traditional classroom settings. Some participants requested payment before they would engage in the projects, adding another layer of difficulty. Time limitations and lack of funding were also frequently mentioned. In some regions, war outbreaks posed severe obstacles to participation. Lastly, not having a Windows operating system for the asteroid search was a technical challenge for some teams.

### 4.2 Challenges and Solutions

The PACS e-Lab management has proposed several solutions to mitigate the challenges faced by participants, as detailed below.

Funding and Resource Constraints: It should be noted that PACS e-Lab operates without external funding. The organization relies on collective resources to execute its projects. Although they have applied for grants in the past, the results have been largely unsuccessful, except for the Astronomy 2024 grant. For this reason, PACS e-Lab management has stated that any of the teams are free to apply for any available grants they qualify for by quoting PACS e-Lab's projects to support their endeavors. Management will not take offense to this. Currently, each participating group is responsible for securing resources to execute the projects in their community. Management supports them by proofreading proposals and donating incentives such as telescopes.

Resource and Support Limitations: While many citizen astronomers participate in and appreciate the projects, there is often a deeper interest in the materialistic aspects of the industry. Sustaining enthusiasm and the opportunity provided by NASA to handle space data can be challenging, as some expect material rewards for their resources, hard work, and time spent. This issue is not unique to Africa but is a global challenge. To address this, PACS e-Lab, in collaboration with its global partners, strives to issue certificates of achievement to successful individuals in all its projects. Additionally, PACS e-Lab has integrated astronomy research writing to train citizen scientists in basic research writing on double stars. These publications will help in their future academic and professional endeavors. PACS e-Lab also rewards a few of its dedicated participants with telescopes so they can develop careers in astronomy outreach in their local communities while continuing to participate in the projects. All these incentives are provided free of charge, thanks to the support of PACS e-Lab's regional and international collaborators.

Geographic and Infrastructural Barriers: In many African countries, the high cost of the internet and unstable power supply make scheduling Zoom meetings very difficult. To resolve this issue, PACS e-Lab and its team of volunteers have made prerecorded tutorials available for all its projects, mostly in English, and are gradually recreating them in Arabic, French, and other local languages spoken in Africa. These recorded tutorials are distributed to the participants, and so far, this approach has mitigated the cost of internet subscriptions that participants would have otherwise exhausted while learning about the projects in the affected regions.

Language Barriers: Although English is the main language of engagement in PACS e-Lab, the management sometimes requests assistance from participants who understand English to train new participants who speak the same local language as them. Most of these sessions are recorded for future reference.

Lack of Interest in Astronomy: Like many developing countries, Africa has a large population that is misinformed about space. Science communicators often face criticism. To recruit participants who might have a passion for space and astronomy but lack opportunities, PACS e-Lab sought collaboration with international organizations like the International Astronomy and Astrophysics Competition (IAAC), which has a database of over 900 African students. To demonstrate their commitment and establish a strong relationship, PACS e-Lab paid the participation fee for 10 African students from 10 different countries in Africa. This gesture attracted the attention of IAAC, leading to PACS e-Lab becoming one of their partners. IAAC agreed to use their platform to promote PACS e-Lab projects among their African participants.

Lack of Central Workstation: Although the projects have been largely successful, PACS e-Lab lacks a central workstation to organize and execute tasks. This is also due to a lack of funding. This remains one of the issues for which PACS e-Lab has not found a solution, and is looking forward to a brighter day.

**4.3 PACS e-:Lab's proposed contributions to the Development of Africa**

PACS e-Lab projects play a significant role in contributing to the development of Africa. These developments will be discussed in the light of the United Nations Sustainable Development Goals (SDGs):

Quality Education (SDG 4): PACS e-Lab projects significantly enhance the quality of STEM education in Africa through the hands-on activities in asteroid search, exoplanet research, Astrophoto visual development, double star research, and more. Most of the partner institutions across the continent have successfully integrated these projects into their curriculum. Some participants have discovered new asteroids and, in the coming years, will have the opportunity to name them. Others have participated in research that has already been published in peer-reviewed journals and are preparing to present at conferences. These endeavors help integrate African researchers into the global scientific community.

Furthermore, the hands-on approach to learning through citizen science projects simplifies complex astronomical concepts, improving comprehension and retention. Participants gain invaluable experience using sophisticated web telescopes like the MicroObservatory (MOBS), Las Cumbres Observatory (LCO), and Slooh, the same observational tools used by professional astronomers. This practical exposure to advanced tools not only demystifies scientific research but also ignites a passion for continuous learning and discovery in the field of astronomy.

Gender Inequality (SDG 5): PACS e-Lab projects significantly address Gender Inequality in STEM fields by actively engaging and empowering African women in space research. Through the online platform, women and girls gain equitable access to STEM education, breaking down barriers that have historically limited their participation in space science and astronomy. This empowerment through skill development not only contributes to narrowing the gender gap in space research participation but also enhances the confidence and self-efficacy of women, enabling them to pursue leadership roles and opportunities for advancement in STEM fields.

Decent Work and Economic Growth (SDG 8): PACS e-Lab initiatives are aligned to foster economic growth through the promotion of decent work, particularly in the realm of STEM. By offering training in research and data analysis, the projects equip participants with a robust skill set encompassing data collection, analysis, and interpretation—skills that are indispensable in today's data-driven world.

This empowerment opens up a multitude of career opportunities in fields such as astronomy, astrophysics, and space science, potentially sparking a lifelong interest in these areas among young African citizen scientists. Moreover, the practical experience gained from the citizen science projects lays a solid foundation for professions in science education, public outreach, and science communication.
These roles are critical in enhancing STEM literacy and ensuring that scientific knowledge is accessible to all. Ultimately, the skills imparted by the projects are not only relevant but also adaptable, providing a versatile toolkit that the participants can apply in a variety of scientific and technical disciplines, thereby contributing to sustainable economic development and the creation of quality jobs in STEM fields.

Industry, Innovation, and Infrastructure (SDG 9): PACS e-Lab's initiative fosters innovation and bolsters the scientific infrastructure in Africa. PACS e-Lab collaborates with esteemed international organizations and makes use of their robotic facilities such as the Las Cumbres Observatory, Slooh telescopes, the MicroObservatory, NASA, etc. Through their training programs and by utilizing their advanced technologies and software tools, the participants are enabled to contribute valuable data to planetary defense and overall astronomical research. These PACS e-Lab activities are instrumental in nurturing a scientific ecosystem within the African region. The activities have the potential to lead to the creation of research facilities, space technology hubs, and a skilled workforce in the field of space science and astronomy, thereby contributing to the overall economic development of the African continent.

Reduced Inequality (SDG 10): Hands-on skills in space science and astronomy are frequently scarce in developing regions, including Africa. PACS e-Lab is committed to addressing this disparity by offering an inclusive online platform that welcomes participation irrespective of geographical location or socioeconomic status. Having reached over 600 individuals from as many as 45 African countries—and with plans for further expansion—the projects are pivotal in narrowing the divide in access to advanced scientific knowledge and skills. This effort not only mitigates inequality within the continent but also plays a crucial role in integrating African researchers into the international astronomy community.

Building Strong Institutions (SDG 16): PACS e-Lab plays a pivotal role in reinforcing institutional strength. By partnering with a wide range of institutions—including national space agencies, astronomy organizations, university physics, and other STEM departments, as well as secondary schools throughout Africa—PACS e-Lab enhances institutional capabilities. These collaborations facilitate practical training, hands-on experiences, and the exchange of knowledge within the African space science and astronomy sectors. By bolstering the educational and outreach initiatives of these entities, the project cultivates a spirit of scientific exploration and cooperation, thereby contributing to the achievement of SDG 16 and the establishment of robust institutional structures across the continent.

Partnerships for the Goals (SDG 17)**:** At the heart of PACS e-Lab's mission is the commitment to building a community of viable African citizen scientists and space enthusiasts. This initiative is a testament to the power of collaboration, which is essential for the advancement of astronomy and citizen science across the continent. PACS e-Lab exemplifies this by forging strategic partnerships with leading international organizations such as the International Astronomical Search Collaboration (IASC), Las Cumbres Observatory (LCO), Institute for Student Astronomical Research, and NASA Exoplanet Watch, as well as a host of African Space and astronomy institutions and organizations. These alliances underscore the vital role of international cooperation in propelling scientific research and education forward. Such partnerships are instrumental in facilitating the exchange of knowledge and building scientific capacity within Africa, thereby nurturing a collaborative ecosystem that is essential for achieving the goals of SDG 17.

### 4.4 Future Directions and Goals

Moving forward, PACS e-Lab will continue to expand its existing projects, particularly double star research, Astrophoto Visual Development, and Exoplanet & photometry, which are not yet widespread as the asteroid search among citizen astronomers. There are also new projects in development called ARISS contact with astronauts aboard the International Space Station. This opportunity will enable the African citizen astronomers to interact with astronauts in the orbiting laboratory by asking questions and receiving replies, just as students from advanced countries do.

Another project called Space-Water-Environment Nexus research and educational non-profitable e-center (SWEN) by Dr. Sherine Ahmed El Baradei affiliated with the German University of Cairo, aims to educate all people whether specialized or enthusiasts from different educational backgrounds. This e-center is based on the idea that space, water, and the environment are strongly interrelated. Water mining on planets; wastewater treatment onboard spacecraft; global warming effects on space debris are main concerns of many space agencies. Managing water resources on earth with aid of satellites is a great breakthrough which will aid in solving the water scarcity problem. The center will provide webinars, seminars, and educational materials.

PACS e-Lab will pursue new, innovative 21st-century astronomy projects to engage the African citizen science community and enhance their research experiences. Additionally, PACS e-Lab plans to organize in-person workshops to disseminate the projects and collaborate with African regional astronomy bodies to organize award events celebrating the achievements of African citizen scientists.

PACS e-Lab plans to continue supporting its African citizen astronomers who are interested in astronomy outreach by providing telescopes, books, and other incentives to enhance the astronomy outreach experience in their communities.

For the few African countries that have not yet reached, PACS e-Lab plans to refine its methodology to identify deserving groups that might appreciate and adopt these projects.

Depending on the availability of funds, PACS e-Lab management hopes to acquire temporary office space to facilitate its work in an organized environment. In the long term, PACS e-Lab management plans to upgrade its location by building an observatory that will serve as a center for astronomy research, education, and outreach endeavors in Africa.

## 5. 0 Conclusion

This paper describes the Pan-African Citizen Science e-Lab (PACS e-Lab) as an online Laboratory platform engaging Africans in hands-on astronomy and space sciences to promote astronomy research, education, and outreach in the continent. PACS e-Lab aims to import some of the captivating hands-on activities in astronomy and space science from across the developed worlds into Africa. This effort helps to bridge the knowledge gap between professionals and non-professionals with an interest in the field between the developed world and Africa.

One of the objectives of the PACS e-Lab is the dissemination of citizen science and soft astronomy research projects, including Asteroid Search, Astro-photo visual development (APVD), Research Writing, Exoplanet and Photometry, the ARISS event, etc across the entire continent. Participants are provided with adequate training to engage in these meaningful and effective projects, enabling them to contribute to space research and exploration. Over 30 asteroids discovered by them have been confirmed by the Minor Planet Center at Harvard. Some participants can utilize web telescopes for the study of double stars and exoplanets and have published their first research papers. Others have acquired skills in astrophotography, enabling them to process astronomical data to produce beautiful visuals. Additionally, tertiary-level teachers and researchers are beginning to adopt these projects and incorporate them into their classroom activities at the bachelor's and master's levels.

Another important aim of PACS e-Lab, which will enable it to fulfill its objectives, is establishing partnerships at both the international and regional levels. International organizations such as the International Astronomical Search Collaboration, Las Cumbres Observatory, Micro Observatory, and NASA Exoplanet Watch provide astronomical data and robotic facilities for remote observation. Meanwhile, regional collaborators have access to deserving groups that could benefit from these initiatives. According to survey data, PACS e-Lab currently collaborates with 15 entities classified as national space agencies, national astronomical societies, and national observatories. Additionally, they work with 25 university groups, 30 private organizations, and secondary schools to facilitate their objectives.

The various PACS e-Lab projects and the skills acquired in the process contribute immensely to the development of Africa, aligning with the United Nations Sustainable Development Goals (UN-SDGs).

The integration of these projects into the curriculum across various institutions of learning in Africa, the hands-on nature of the projects, student-led publications, and the use of web telescopes for education and research all contribute to Quality Education (SDG 4).

PACS e-Lab's engagement of women and girls in hands-on astronomy activities, who make up approximately 38%, and efforts to increase their number, break down barriers limiting their participation in STEM fields, addressing Gender Inequality (SDG 5).

The skill set encompassing data collection using web telescopes, analysis using software like Astrometrica, EXOTIC, and AstroImageJ, and interpretation—skills crucial in fields such as astronomy, space science, STEM education, public outreach, and science communication, and transferable to other scientific and technical disciplines—contributes to Decent Work and Economic Growth (SDG 8).

PACS e-Lab's collaboration with web telescope operators and NASA, and utilization of their advanced technologies, nurture a scientific ecosystem within the African region and could trigger the creation of research facilities, space technology hubs, and a skilled STEM workforce, contributing to Industry, Innovation, and Infrastructure (SDG 9). From the survey, the majority of participants stated that PACS e-Lab projects were their first hands-on endeavors in astronomy and space science activities.

Moreover, having disseminated the projects across 45 out of the 54 countries in Africa, reaching institutions of learning in underserved countries, contributes to Reduced Inequality (SDG 10).

Various entities, including national space agencies, astronomy organizations, university physics, and other STEM departments, and secondary schools partnering with PACS e-Lab, strengthen their program deliverables, resonating with Building Strong Institutions (SDG 16).

PACS e-Lab activities in Africa over the years, as deduced from this literature, truly demonstrate that it serves as a bridge bringing hands-on activities from the global north to Africa, bridging the knowledge gap in astronomy and space science practices for sustainable development, aligning with Partnerships for the Goals (SDG 17).

Judging from the fact that since the activities of PACS e-Lab began in late 2020 until the second quarter of 2024, the time of this literature, its scientific endeavors through citizen science have gained widespread acceptance. This acceptance spans national space agencies, tertiary institutions, national observatories, astronomy organizations and clubs, and secondary schools. Up to 70 groups of varying sizes, cutting across 45 out of the 54 countries in Africa, have been engaged. Additionally, despite challenges stemming from a lack of electricity, computers, funding, strong internet services, and the high cost of internet access, we can conclude that citizen science is a powerful tool for spreading STEM knowledge and skills across developing regions like Africa.

Looking at the enormous progress made by PACS e-Lab in making hands-on astronomy and space science accessible in Africa, it is important to call for increased support and collaboration. The challenges faced by African citizen scientists, such as the lack of computers, reliable internet services, and adequate funding, highlight the urgent need for resource mobilization.

PACS e-Lab management urges government educational ministries across African countries, international organizations like the International Astronomical Union, NASA, JAXA, and private sector stakeholders to invest in these infrastructures, ensuring that citizen science can continue to thrive and expand across Africa. The provision of these necessary tools can empower more individuals to participate in scientific research in Africa.

PACS e-Lab extends its partnerships to more organizations across the continent to join forces in engaging the African public in its projects.

**Data Availability Statement**

All data used for this study were obtained through a survey. The survey data were summarized in a Google Sheet, which is included in this article. The original survey data will be made available upon request.

**Acknowledgment**

The PACS e-Lab teams are grateful to their international partners: International Astronomical Search Collaboration (IASC), NASA Exoplanet Watch, Las Cumbres Observatory, and the MicroObservatory, for their inclusion spirit and supply of datasets for asteroid searches, the Exoplanet Citizen Science project, and the opportunity to conduct observations on exoplanets with their remote telescope facilities, respectively.

PACS e-Lab also thanks Dr. Rachel Freed and Ms. Kalee Tock for training their team in double-star research, which has been successfully incorporated into PACS e-Lab.

The PACS e-Lab management additionally extends its gratitude to the Jean Pierre Grootaerd Telescope for All Programs for donating telescopes to some of PACS e-Lab's citizen scientists

PACS e-Lab management also thanks Dr. Charles Takalani, the head of the Secretariat for the African Astronomical Society, for always being there for PACS e-Lab, assisting in various ways such as providing supporting letters and signatures for PACS e-Lab's international collaboration whenever needed. Many thanks also to the AfAS and the IAU-GA 2024 Cape Town team for the grant (Number: 14) and telescope donations to support the efforts of PACS e-Lab. The grant partially supported the projects.

PACS e-Lab management is grateful to Ezeakunne Chidozie (PhD candidate) for donating to PACS e-Lab to support its efforts.

PACS e-Lab extends warm appreciation to all its volunteers and team leadership for their efforts.

**Conflict of interest**

The authors declare that there are no potential conflicts of interest in this publication.

**Ethical statement**

All the participants were informed about the publication and they granted their consent to the work.

**Authors' Contribution**

All authors have been involved in the PACS e-Lab project. Miracle Chibuzor Marcel wrote the drafts of the paper. The rest of the authors are the team leaders and members of various units. They filled out the survey that provided data for this study.